# Scaling of Volumetric Data in Model Systems Based on the Lennard-Jones Potential

A. Grzybowski, K. Koperwas, and M. Paluch

*Institute of Physics, University of Silesia, Uniwersytecka 4, 40-007 Katowice, Poland*

The crucial problem for better understanding the nature of glass transition and related relaxation phenomena is to find proper interrelations between molecular dynamics and thermodynamics of viscous systems. To gain this aim the recently observed density scaling of viscous liquid dynamics has been very intensively and successfully studied for last years. However, previous attempts at related scaling of volumetric data yielded results inconsistent with those found from the density scaling of molecular dynamics. In this Letter, we show that volumetric data obtained from simulations in simple molecular models based on the Lennard-Jones (LJ) potential, such as Kob-Andersen binary liquids and the Lewis-Wahnström o-terphenyl model, can be scaled by using the same value of the exponent, which scales dynamic quantities and is directly related to the exponent of the repulsive inverse power law that underlies short-range approximations of the LJ potential.

For recent years, theoretical investigations of the glass transition have taken a promising direction [1] supported by many observations [2 - 10] based on experimental data, mainly for supercooled van der Waals liquids and polymer melts, showing that some dynamic quantities, such as structural (or segmental in the case of polymer melts) relaxation time τ and viscosity η, can be plotted onto one master curve according to a function

$$\log_{10} x = f\left(T^{-1}\upsilon^{-\gamma}\right) \quad (1)$$

where $x$ = τ or η (and also the diffusion coefficient $D$), $T$ is the system temperature, υ is the system specific volume, and γ is a material constant independent of thermodynamic conditions near the glass transition. The found validity of the phenomenological description of experimental data by Eq. (1), commonly called as 'thermodynamic scaling' or 'power-law density scaling', has stimulated a lot of efforts into a better understanding of molecular determinants of this scaling and its relation to macroscopic thermodynamic properties of viscous systems. The most popular approach [11 - 18] to this study, strongly exploiting molecular dynamics (MD) simulations, relies on the generalized Lennard-Jones potential [19]

$$U_{LJ}(r) = 4\varepsilon\left[(\sigma/r)^m - (\sigma/r)^n\right] \quad (2)$$

and its effective approximation for small intermolecular distances

$$U_{eff}(r) = 4\varepsilon(\sigma/r)^{m_{IPL}} - A_t \ , \quad (3)$$

which is believed to be responsible for the thermodynamic scaling. Attractive forces in Eq. (3) are reflected by some constant (or linear) small background $A_t$. The main part of $U_{eff}$ is constituted by a repulsive inverse power law (IPL) term $U_{IPL}$ proportional to $r^{-m_{IPL}}$, where one usually assumes that $m_{IPL} = 3\gamma_{IPL} > m$. According to many MD simulations [13 - 18] based on Eq. (2), the parameter $\gamma_{IPL}$ established for simulation data can be identified with the exponent γ that enables to scale values of τ, η or D (obtained from the same simulation data) in terms of Eq. (1). Very recently, Pedersen et al. [20] have shown that the isochoric heat capacity calculated from simulations of the Kob-Andersen binary Lennard-Jones (KABLJ) liquid [21] can be scaled also by using the same value of $\gamma$.

However, related investigations of the scaling of volumetric data have not provided until recently such spectacular results. Alba-Simionesco *et al.* [5] have observed no scaling of plots of $pv/T$ vs $T^{-1/\gamma}v^{-1}$ after using pressure-volume-temperature ($pvT$) experimental data with the exponent $\gamma$ determined from the scaling of structural relaxation times according to Eq. (1). Roland et al. [22] have noted that an equation of state (EOS) based only on a pure IPL, $Ar^{-3\gamma}$, is inaccurate due to excluding attractive interactions, and they proposed only the pressure dependence of the glass transition temperature, which involves the scaling exponent $\gamma$, by adapting Simon's [23] empirical relation $p(T)$, which has been rationalized [19] also by the IPL potential, to the thermodynamic scaling requirements. Later, Schrøder et al. [18] have even argued that the equation of state based only on the pure IPL does not obey the inverse power-law density scaling. Realizing that attractive forces cannot be omitted, Alba-Simionesco and Tarjus [12] have assumed that attractive interactions are represented by a uniform background added to the repulsive part of the Lennard-Jones potential, and then they have tried to scale $pvT$ data by plotting values of the expression $pv/T + a/(vT)$ vs $T^{-1/\gamma}v^{-1}$, where the term $a/(vT)$ follows from the mean-field attractive forces. As a result they achieved a satisfactory scaling of volumetric data of o-terphenyl (OTP) for the value of $\gamma$ larger than that determined from the scaling of viscosities or structural relaxation times.

In this Letter, we answer the important question for our proper understanding of interrelations between dynamics and thermodynamics of viscous systems: *Can one scale dynamic and volumetric data with the same value of the exponent $\gamma$ clearly related to intermolecular interactions which are relevant to viscous systems?*

To inspect directly intermolecular interactions we focus on investigations of the KABLJ liquid [21], which is the well-known model of supercooled liquids also commonly exploited to study the power-law scaling of molecular dynamics in viscous systems [13-18]. We used the RUMD package [24] to perform MD simulations (Section I in Ref. 25). According to theoretical and simulation investigations performed by Dyre's group, the viscous KABLJ system belongs to *strongly correlating* liquids that are characterized by a strong linear *WU* correlation [13,14,15] between the system virial $W$ and the system potential energy $U$: $\Delta W(t) \cong \gamma \Delta U(t)$, where $\Delta U(t) = U(t) - <U>$ and $\Delta W(t) = W(t) - <W>$ are respectively deviations of the instantaneous values $U(t)$ and $W(t)$ from their thermal averages $<U>$ and $<W>$. We have analyzed the obtained values of the *WU* correlation coefficient $R = <\Delta W \Delta U> / \sqrt{<(\Delta W)^2><(\Delta U)^2>}$ and the *WU* dependence slope $\gamma \equiv \sqrt{<(\Delta W)^2>/<(\Delta U)^2>}$ for each simulation state $(T^*, \rho^*)$, finding that results of our simulations meet the criterion of the strong *WU* correlation ($R$>0.9) and the slopes $\gamma$ are temperature- and density-dependent. Since the variations of the *WU* slope are more sensitive to changes in density we can determine an average value of $\gamma$ over different temperatures for a given density (FIG. 1 in Ref. 25). This already known result of LJ-based MD simulations [14] shows that the density scaling is imperfect for such systems in the exact sense, but some effective mean value of the scaling exponent $\gamma$ can be estimated over different temperatures and densities according to the general formula for thermodynamic scaling (Eq. (1)). The most convenient way to do that is to use a model that enable us to determine the effective value of $\gamma$ to a good approximation. To date the best one [1] is the temperature-volume version [26] of the Avramov entropic model [27] shown herein for structural relaxation times

$$\log_{10}\tau = \log_{10}\tau_0 + \left(\frac{A}{T\upsilon^{\gamma}}\right)^F \log_{10} e \tag{4}$$

where $\tau_0$, $A$, $F$, $\gamma$ are fitting parameters shared for a whole analyzed set of temperatures $T$ and specific volumes $\upsilon$. In the case of simulation data, we can replace the specific volume $\upsilon$ in Eq. (4) with the average dimensionless volume per molecule $\upsilon^*=V^*/N$, which is fixed for each NVT ensemble, and also $T$ and $\tau$ with their dimensionless counterparts $T^*$ and $\tau^*$.

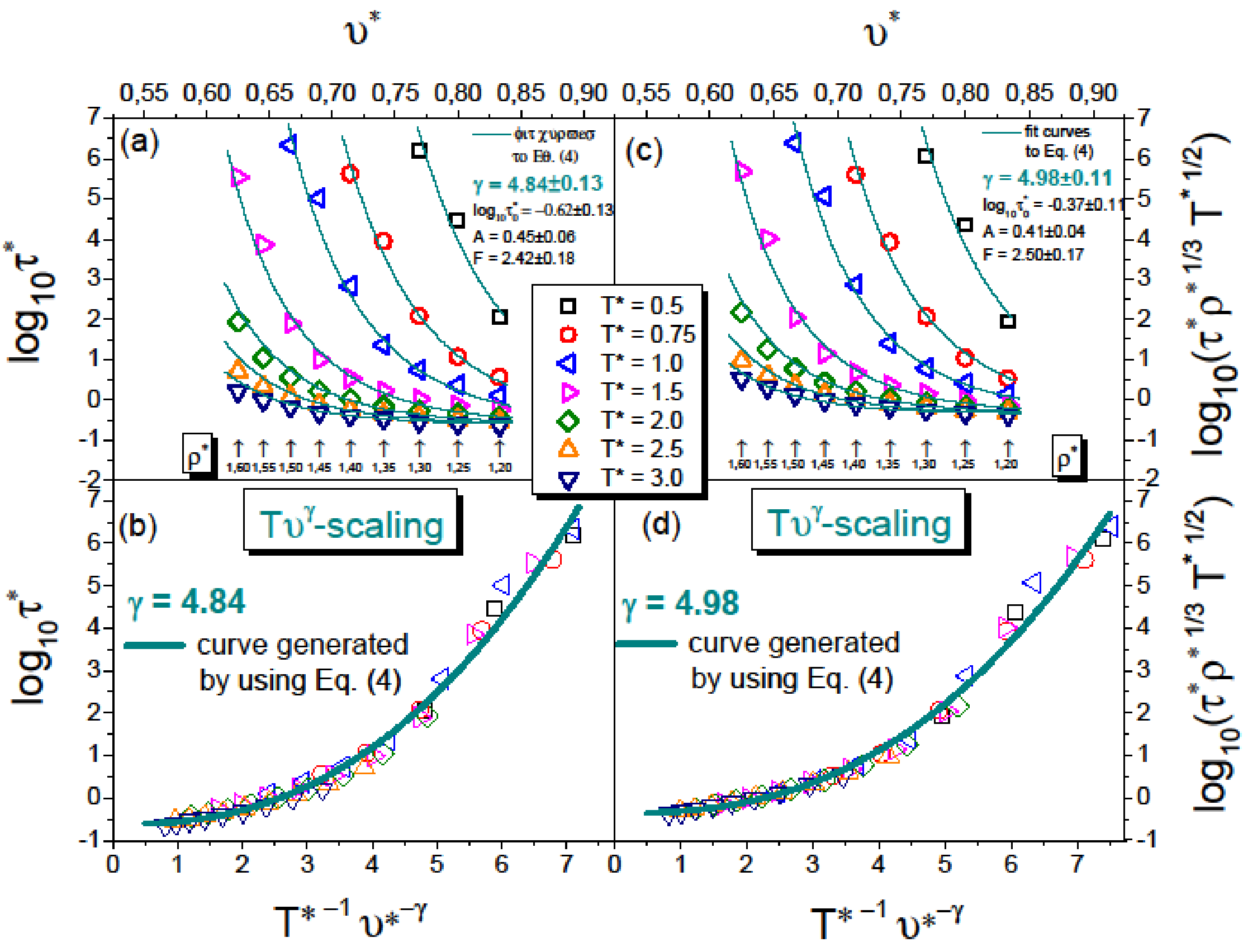

FIG. 1

Plots of the volume dependences of (a) isothermal dimensionless relaxation times $\tau^*$ fitted to Eq. (4) and (c) their representations in reduced units also fitted to Eq. (4). The thermodynamic scaling with the fitted value of the scaling exponent $\gamma$ is shown in the panels (b) and (d) for $\tau^*$ and $\tau^*$ in reduced units, respectively.

In our analysis, the structural relaxation times $\tau^*$ are calculated from simulation data typically [20], i.e., by using [21] incoherent intermediate self-scattering functions ($\tau^* = t^*$ if $F_S(q^*, t^*) = e^{-1}$) determined at the wave vector $q^*$ of the first peak of the AA structure factor for each simulation state ($T^*$,$\rho^*$) separately. The isothermal volume dependences of $\tau^*$ with its fit to Eq. (4) are shown in FIG. 1(a). When we apply the found fitting value of the scaling exponent $\gamma$=4.84±0.13 to scale $\tau^*$ vs $T^{*-1}\upsilon^{*-\gamma}$ we can notice from the inset to FIG. 1(b) that the power-law density scaling of structural relaxation times in terms of Eq. (1) is rather roughly satisfactory. As already mentioned it corresponds to the density-dependent WU correlation (FIG. 1 in Ref. 25). Nevertheless, strongly correlating liquids exhibit a hidden scale invariance [28], which enables to better reveal the scaling behavior [18,20] by using some reduced units, e.g., such units for which NVE and NVT Newtonian dynamics are isomorph invariant [29]. To meet this requirement we multiply the obtained structural relaxation times $\tau^*$ for our KABLJ system by $\rho^{*1/3}T^{*1/2}$ (FIG. 1(c)). Then, the power-law density scaling of $\tau^*$ in the reduced units is improved (FIG. 1(d)), but this effect is not observed for low temperature isotherms at higher densities. The found mean value of the scaling exponent $\gamma$ = 4.98±0.11 for $\tau^*$ in the reduced units, which is only slightly different from that obtained for our KABLJ system without using these units, enables to scale molecular dynamics of the viscous KABLJ liquid in the relatively wide temperature-density range to a quite good approximation.

As mentioned the scaling exponent $\gamma$ is commonly identified with the exponent $\gamma_{IPL}$ = $m_{IPL}/3$, where $m_{IPL}$ is the exponent of the repulsive IPL term of the effective intermolecular potential $U_{eff}$ given by Eq. (3), which is the short-range approximation of the LJ potential. Assuming that the effective intermolecular potential can be expressed just by Eq. (3), we have previously derived [30,31,32] a very important equation of state

$$\left(\frac{\upsilon_0}{\upsilon}\right)^{\gamma_{EOS}} = 1 + \frac{\gamma_{EOS}}{B_T^{conf}(p_0^{conf})}\left(p^{conf} - p_0^{conf}\right), \tag{5}$$

which is presented herein for the configurational pressure, $p^{conf} \equiv p - RT/(M\upsilon) \cong <W>/V$ , where $R$ is the gas constant, $M$ is molar mass, υ and V are the system volume and specific volume respectively, and $B_T^{conf}(p_0^{conf})$ is the isothermal configurational bulk modulus in a chosen reference state [33]. This EOS can be used in the thermodynamic scaling regime, expecting that $\gamma_{EOS} \approx \gamma_{IPL}$, because Eq. (5) is formulated for the low compressibility region from the system average virial $<W> = -(1/3)\left\langle \sum_i \mathbf{r}_i \cdot \nabla_i U \right\rangle$ on the assumption that $U$=$U_{eff}$. Since $U_{eff}$ given by Eq. (3) is mainly based on the repulsive $U_{IPL}$, then $<W> \sim r^{-3\gamma_{IPL}} \sim \upsilon^{-\gamma_{IPL}}$. Thus, the parameter $\gamma_{EOS}$ in Eq. (5) should be identified with $\gamma_{IPL}$ similarly to the slope $\gamma$ of the *WU* correlation. It means that one should be able to determine $\gamma_{IPL}$ by fitting volumetric data to Eq. (5). Moreover, Eq. (5) implies some kind of power-law scaling of volumetric data [31,32] if the value of the parameter $\gamma_{EOS}$ does not depend on thermodynamic conditions. Until recently, we have been exploiting Eq. (5) to fit and scale very successfully only experimental PVT data for supercooled liquids and polymer melts. However, the expected correspondence $\gamma_{EOS} \approx \gamma_{IPL}$ has not been undoubtedly proven yet [30 - 36]. Herein, for the first time, the EOS given by Eq. (5) as well as another EOS [31,32,36,37] that describes experimental data similarly to Eq. (5) are verified by using $p\upsilon T$ data taken from MD simulations of the model viscous liquids.

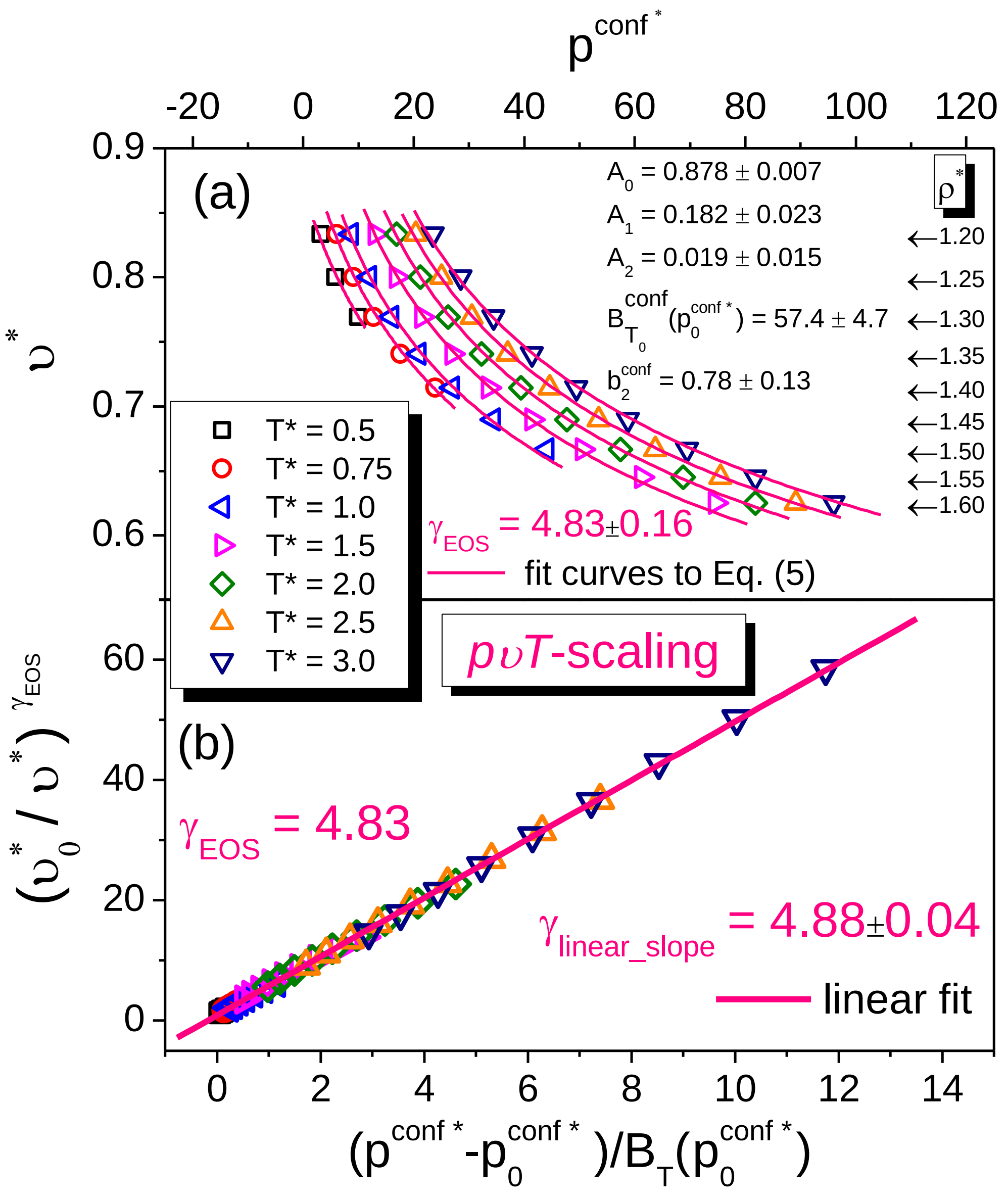

FIG. 2

(a) Plot of isothermal average dimensionless volumes per molecule $\upsilon^*$ vs dimensionless configurational pressures $p^{conf^*}$ and their fit to Eq. (5). (b) Scaling of volumetric data with the scaling exponent $\gamma_{EOS}$ found from the fitting to Eq. (5). A linear fit of $p^{conf}\upsilon T$ data, which is independent of Eq. (5), demonstrates the high quality of $p^{conf}\upsilon T$ -scaling.

In the case of simulation data, we can replace the specific volume υ in Eq (5) with the average dimensionless volume per molecule $\upsilon^*=V^*/N$ and the quantities $T$, $p^{conf}$, $B_T^{conf}$ with their dimensionless counterparts $T^*$, $p^{conf*}, B_T^{conf*}$. We need to note that this EOS is independent of the used units due to its quotient terms. Thus, the reduced unit of pressure is not required in contrast to that applied to structural relaxation times. Then, we can fit volumetric data directly obtained from MD simulations of the KABLJ liquid to Eq. (5). As a result (FIG. 2(a)), we find a good quality fit of $p^{conf}\upsilon T$ data with the value $\gamma_{EOS}=4.83\pm0.16$, which is very close to the value of the exponent γ that scales structural relaxation times vs $T^{-1}\upsilon^{-\gamma}$. This meaningful finding $\gamma_{EOS} \approx \gamma$, which is also maintained for structural relaxation time in reduced units, leads to the expected but until recently unobvious conclusion that beside the slope γ of the *WU* correlation and the scaling exponent γ of structural relaxation times also the exponent $\gamma_{EOS}$ in the EOS given by Eq. (5) can be identified with the IPL exponent $\gamma_{IPL}$ at least for the viscous KABLJ liquid.

Next, we perform the scaling of volumetric data by using the fitted value of the exponent $\gamma_{EOS}$. As can be noted Eq. (5) predicts a linear scaling of the pure power law for the reduced specific volume ($\upsilon_0/\upsilon$) or reduced density ($\rho/\rho_0$) vs the reduced change in the configurational pressure $\left(p^{conf} - p_0^{conf}\right)/ B_T^{conf}(p_0^{conf})$. It is shown in FIG. 2(b) that this scaling of volumetric simulation data for the viscous KABLJ liquid with $\gamma_{EOS}=4.83$, previously determined by fitting $p^{conf}\upsilon T$ data to Eq. (5), is characterized by a high quality which is better than that achieved for the power-law scaling of structural relaxation times. An analogous scaling of the other tested EOS has been also positively verified (Section III in Ref. 25). As a consequence the found values of $\gamma_{EOS}$ are numerically close to each other and also to the values of the scaling exponent γ estimated from structural relaxation times calculated from the simulation data with using the reduced units of τ* as well as without these units.

The very successful results obtained for the KABLJ model strongly encouraged us to verify whether this is a unique case or not. Therefore, we have continued investigating the problem and applied Eq. (5) to fit simulation volumetric data for the Lewis-Wahnström (LW) model [38] of OTP collected by Dyre's group. As a result, we have obtained $\gamma_{EOS}=8.20\pm0.23$ for this model of OTP, whereas Dyre's group [14,18] and Tarjus et al. [39] earlier found also γ≈8 as both the slope of the *WU* correlation and the value of the exponent that enables the power-law density scaling of dynamic quantities (e.g. diffusivity) for the same model of OTP.

In conclusion, we have found that volumetric data for the simple model systems such as the Kob-Andersen binary Lennard-Jones liquid and the Lewis–Wahnström three-site Lennard-Jones model for ortho-terphenyl can be perfectly scaled according to the EOS given by Eq. (5) and also to the EOS described in Section III in Ref. 25. This scaling is achieved for both the equations of state with the values of the scaling exponent $\gamma_{EOS}$ which correspond very well to the value of the exponent γ for which dynamic quantities such as structural relaxation time, viscosity, and diffusivity obey the power-law density scaling vs $T^{-1}\upsilon^{-\gamma}$ (also called thermodynamic scaling). The obtained numerical correspondence between $\gamma_{EOS}$ and γ as well as the assumption of the effective potential $U_{eff}$ mainly based on the repulsive inverse power law, which has been made to derive Eq. (5), allow us to identify the exponent $\gamma_{EOS}$ with $\gamma_{IPL}$=

$m_{IPL}/3$, where $m_{IPL}$ is the exponent of the repulsive part of the short range approximation (Eq. (3)) of the Lennard-Jones potential (Eq. (2)). The tested herein representative simple models show that the obtained meaningful result should be valid for each simple model based on the Lennard-Jones potential within the range of the strong *WU* correlation. However, further investigations are required for real materials near the glass transition, for which we have previously found [30 - 36] that $\gamma_{EOS} > \gamma$.

The authors are thankful for the financial support received for the research project (contract no. TEAM/2008-1/6), which is operated within the Foundation for Polish Science Team Programme co-financed by the EU European Regional Development Fund. The authors are deeply grateful to Dr. Ulf Pedersen and Prof. Jeppe Dyre from Roskilde University in Denmark for making available volumetric data from their MD simulations of the LW OTP model.

## Supplemental Material

### I. Details of Our MD Simulations of the KABLJ Liquid

By using the RUMD package [24], we have performed MD simulations of the 12-6 KABLJ liquid composed of $N$=1000 species ($N_A$=800 of the type $A$ and $N_B$=200 of the type $B$ with their masses $M_A$=$M_B$) interacting via the LJ potential (Eq. (2)) with its typical cuttoff 2.5$\sigma_{AA}$ and parameters: $m$=12, $n$=6, , $\varepsilon_{BB}/\varepsilon_{AA}=0.5$, $\varepsilon_{AB}/\varepsilon_{AA}=1.5$, $\sigma_{BB}/\sigma_{AA}=0.88$, and $\sigma_{AB}/\sigma_{AA}=0.8$, where $\sigma_{AA}$=1.0, $\varepsilon_{AA}$=1.0, and $M_A$=1.0, which are units of length, energy, and mass, respectively. The equilibrium NVT simulations have been done for the supercooled region in the relatively wide temperature-density range, $0.5 \le T^* \le 3.0$ and $1.2 \le \rho^* \le 1.6$, where the dimensionless temperature and number density are defined respectively as $T^* = k_B T / \varepsilon_{AA}$ (with the Boltzman constant $k_B$=1.0) and $\rho^* = N / V^* = \rho\sigma_{AA}^3$ with the dimensionless volume $V^* = V / \sigma_{AA}^3$). Each simulation run has taken at least about $10^8$ time steps with the dimensionless time step length $\Delta t^* = 0.001$, where $t^* = t / [\sigma_{AA} (M_A / \varepsilon_{AA})^{0.5}]$.

## II. The Strong WU Correlation in the Simulated KABLJ Liquid

As can be seen in FIG. 1 in this Supplemental Material we have considered only the viscous KABLJ liquid characterized by the strong linear isochoric WU correlation $\Delta W(t) \cong \gamma \Delta U(t)$ between the system virial *W* and the system potential energy *U*, where $\Delta U(t) = U(t) - \langle U \rangle$ and $\Delta W(t) = W(t) - \langle W \rangle$ are respectively deviations of the instantaneous values *U*(*t*) and *W*(*t*) from their thermal averages $\langle U \rangle$ and $\langle W \rangle$. Moreover, we have found that the values of the *WU* correlation coefficient $R = \langle \Delta W \Delta U \rangle / \sqrt{\langle (\Delta W)^2 \rangle \langle (\Delta U)^2 \rangle}$ meet the criterion for the strong *WU* correlation (*R*>0.9) for each simulation state ($T^*$,$\rho^*$).

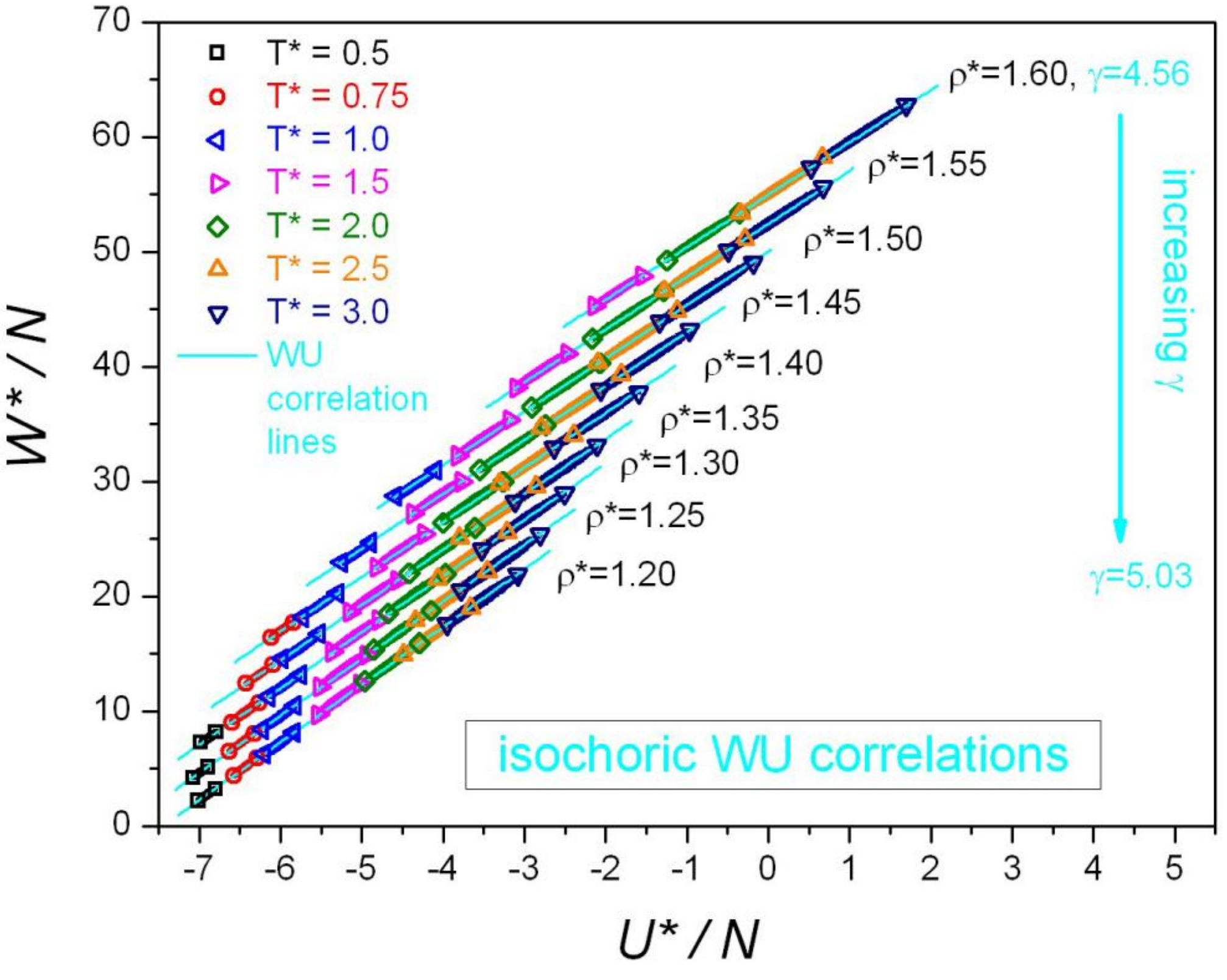

FIG. 1

Plot of correlations between instantaneous values of the dimensionless virial $W^*$ per molecule and the potential energy $U^*$ per molecule. Extreme values of $U^*$ at a given ($T^*$,$\rho^*$) are denoted by big symbols, other points ($U^*$,$W^*$) are marked by small (undistinguishable) symbols.

### III. Scaling of the KABLJ Liquid Volumetric Data Similar to That Implied by Eq. (5)

It is worth noting that there is also another EOS, which has some different origin [31,32,36,37] than that of Eq. (5) but it is morphologically very similar to Eq. (5) and can be obtained by replacing all configurational quantities in Eq. (5) with their non-configurational counterparts as follows

$$\left(\frac{\upsilon(T,p_0)}{\upsilon(T,p)}\right)^{\gamma_{EOS}} = 1 + \frac{\gamma_{EOS}}{B_T(p_0)}\left(p - p_0\right) \qquad (1)$$

This EOS is derived [32] from its isothermal precursor [31,36], which follows [37] from some mathematical modification $B_T = -\gamma_{EOS}\left(\partial p / \partial \ln \upsilon^{\gamma_{EOS}}\right)_T$ of the definition of the isothermal bulk modulus. The volume $\upsilon_0$ is typically parametrized by a quadratic temperature function $\upsilon_0(T) = \upsilon(T,p_0) = A_0 + A_1(T-T_0) + A_2(T-T_0)^2$, the isothermal bulk modulus $B_T(p_0) = B_{T_0}(p_0)\exp(-b_2(T-T_0))$, where the fitting parameter $b_2$ is approximately constant independently of the choice of $T_0$ in the supercooled region, and the fixed reference state is defined herein by $T_0^* = 0.5$, which is the lowest considered temperature, zero pressure $p_0^*$.

In the case of experimental *pvT* data, this EOS yields values of $\gamma_{EOS}$, which are in accord with those found from Eq. (5) and enables to scale very well volumetric data. We also observe the same correspondence of these equations of state for simulation volumetric data for the model KABLJ liquid, for which Eq. (1) shown in this Supplemental Material yields the value $\gamma_{EOS}$=4.58±0.19 that scales very well these *pvT* simulation data (FIG. 2 in this Supplemental Material). Thus, the found values of $\gamma_{EOS}$ are numerically close to each other and also to the values of the scaling exponent γ estimated from structural relaxation times calculated from the simulation data with using the reduced units of τ* as well as without these units.

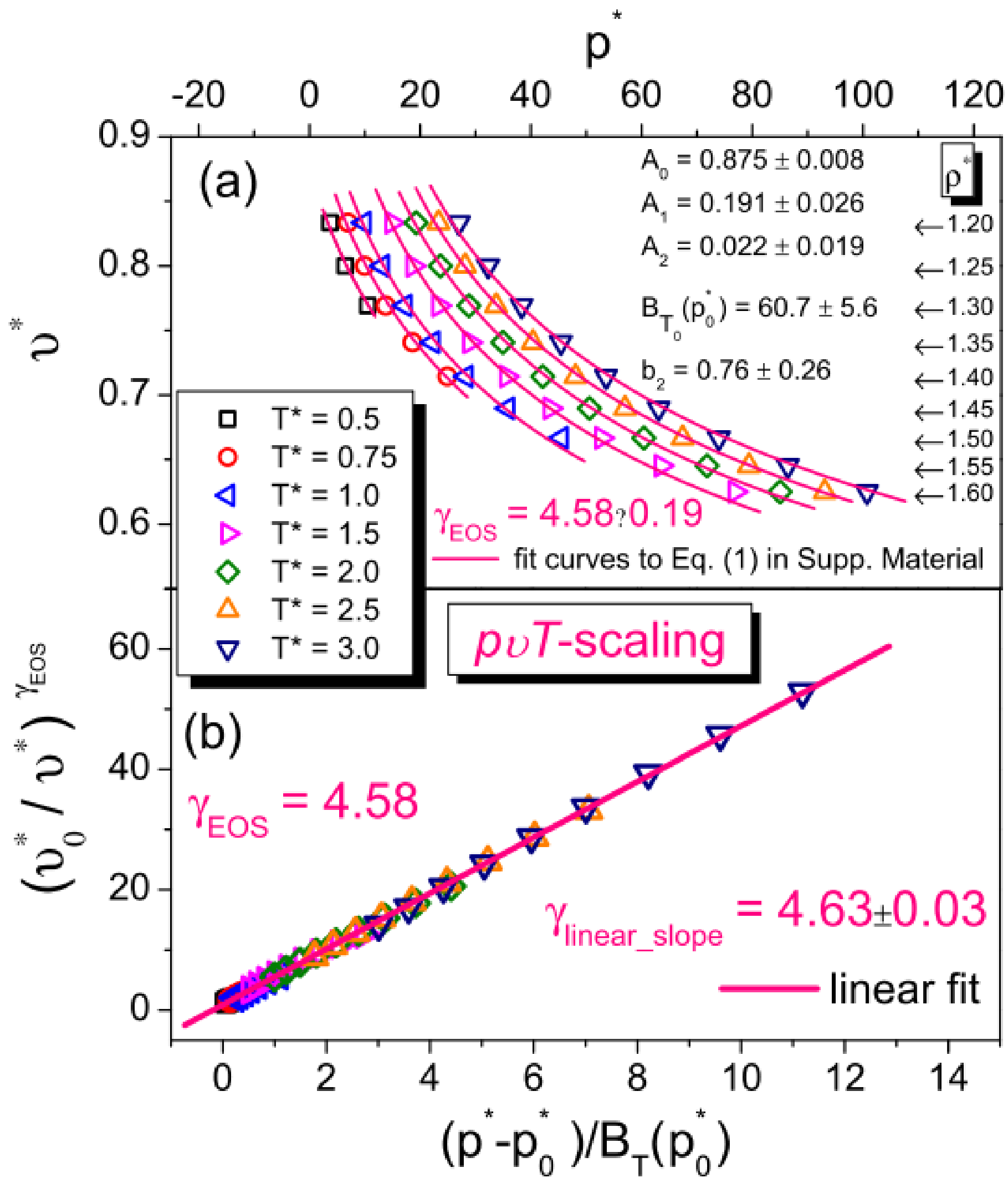

FIG. 2

(a) Plot of isothermal average dimensionless volumes per molecule $\upsilon^*$ vs dimensionless pressures $p^*$ and their fit to Eq. (1) in this Supplemental Material. (b) Scaling of volumetric data with the scaling exponent $\gamma_{EOS}$ found from the fitting to Eq. (1) in this Supplemental Material. A linear fit of $p\upsilon T$ data, which is independent of Eq. (1) in this Supplemental Material, demonstrates the high quality of $p\upsilon T$-scaling.